\pdfoutput=1
\documentclass{article}

\usepackage[T1]{fontenc}
\usepackage{microtype}
\usepackage{graphicx}
\usepackage{booktabs}
\usepackage{multirow}
\usepackage{array}
\usepackage{amsmath}
\usepackage{xcolor}
\usepackage{tikz}
\usetikzlibrary{arrows.meta,positioning,fit,calc,backgrounds}
\usepackage{hyperref}
\usepackage{xurl}
\usepackage{enumitem}
\usepackage{float}
\setlist{nosep,leftmargin=*}

\usepackage[accepted]{mlsys2025}
\makeatletter
\renewcommand{\mlsys@appearing}{Preprint.}
\makeatother
\hypersetup{pdfsubject={}}

\newcommand{\fp}{Firepool}
\newcommand{\x}{$\times$}
\newcommand{\para}[1]{\smallskip\noindent\textbf{#1}\hspace{0.4em}}
\newcommand{\rejected}[1]{\smallskip\noindent\textit{Rejected alternative: #1}\hspace{0.4em}}
\newcommand{\code}[1]{\texttt{\small #1}}
\definecolor{fpblue}{HTML}{0072B2}
\definecolor{fporange}{HTML}{E69F00}
\definecolor{fpgreen}{HTML}{009E73}
\definecolor{fpverm}{HTML}{D55E00}
\definecolor{fpgrey}{HTML}{999999}
\definecolor{fpsky}{HTML}{56B4E9}

\mlsystitlerunning{Firepool: Brokering Warm GPU Capacity for Agentic and RL Workloads}

\begin{document}

\twocolumn[
\mlsystitle{Firepool: Brokering Warm GPU Capacity\texorpdfstring{\\}{ }for Agentic and RL Workloads at LinkedIn}

\begin{mlsysauthorlist}
\mlsysauthor{Ankit Goyal}{li}
\mlsysauthor{Santosh Jha}{li}
\mlsysauthor{Zhuo Zhi}{li}
\mlsysauthor{Youmin Han}{li}
\mlsysauthor{Cong Gu}{li}
\mlsysauthor{Chen Xie}{li}
\mlsysauthor{Animesh Singh}{li}
\end{mlsysauthorlist}
\mlsysaffiliation{li}{LinkedIn}
\mlsyscorrespondingauthor{Ankit Goyal}{\mbox{angoyal@linkedin.com}}

\mlsyskeywords{warm pools, GPU scheduling, Kubernetes, agentic workloads, reinforcement learning, cold start}

\vskip 0.3in

\begin{abstract}
AI agents and reinforcement-learning (RL) trainers run thousands of short, isolated GPU jobs: tool calls, verifiers, and rollouts. A job often finishes in seconds, but starting a GPU pod for it takes 20--35\,s even when the container image is already on the node, and minutes when it is not. A warm pod removes the wait only if it can serve the next job, and keeping pods warm must neither waste GPUs nor leak state between callers. \fp{}, the sandbox and execution broker behind agentic and RL workloads at LinkedIn, meets both requirements. It makes pods interchangeable by building them from a shared base image per hardware profile and binding each job's code, extra dependencies, and cluster topology when the job starts. It keeps them warm cheaply and safely by reusing a pod only for the caller that used it last, staging zero-VRAM standby pods in fractional GPU shares alongside active pods, and capping the GPUs that warm pools may hold via queue-driven budgets. Over a sampled peak period in production, \fp{} serves 94.7\% of 3,206 pod requests from warm or standby capacity. In paired benchmarks with fixed-duration jobs on production H200 nodes, agent and RL loops finish 13\x{} and 10.8\x{} faster than with a fresh pod per job; against a fresh reuse tag per job, warm reuse cuts p90 start latency from 59.1\,s to 3.5\,s. When a job cannot reuse the previous pod, standby promotion shortens the handover wait from 51\,s to 0.94\,s. In local and simulated runs, a keep-alive matched to the loop's pauses keeps the latency benefit while holding about half the capacity of an always-on pool.

\end{abstract}
]

\printAffiliationsAndNotice{\raggedright}

\section{Introduction}
\label{sec:intro}

\begin{figure}[t]
\centering
\includegraphics[width=\columnwidth]{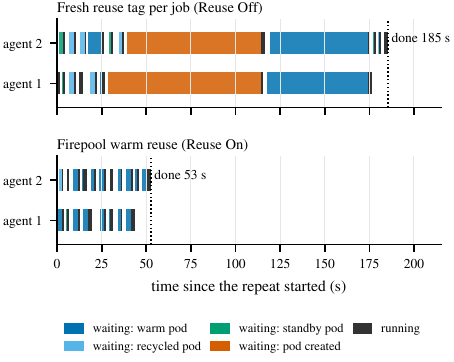}
\caption{One repeat of a two-agent tool loop on a production H200 node under a two-GPU cap. Bars show each job's wait, colored by how its pod was obtained, then its run. Top: a fresh reuse tag per job retires every pod, and jobs that wait for a new pod hold up the agent's next steps; the repeat takes 185\,s. Bottom: \fp{} reuses each agent's warm pod and the repeat finishes in 53\,s. The medians over 10 paired repeats: 186.4\,s and 51.8\,s (\S\ref{sec:eval-critical}).}
\label{fig:timeline}
\end{figure}

A coding agent edits a file, runs a test in a sandbox, reads the result, and tries again. An RL trainer fans out rollouts, scores them, and updates the policy. When a program replaces a person in this loop, steps arrive seconds apart instead of minutes, in bursts, around the clock. On a GPU each step often runs for seconds, but starting a pod for it takes much longer: 20--35\,s on our H200 nodes with the image cached, and 1.3--7.5 minutes on average across our fleet when it is not. Over thousands of short jobs, these waits dominate the loop's running time (Figure~\ref{fig:timeline}).

Common fixes each cover part of the problem. Caching images on nodes removes the network pull but not the 20--35\,s of scheduling, device binding, and platform initialization, and it does not scale when every experiment brings a new image. Persistent workers, such as a Ray cluster or a static rollout pool held for a whole run, remove the wait between steps but hold GPUs idle through every think-time pause and off-peak hour. Batch schedulers such as Kueue and Volcano manage queueing and quotas but do not keep reusable warm pods across independent jobs.

\fp{} is the sandbox broker behind agentic and RL workloads at LinkedIn. Its goal is that no job waits for a pod to be created once capacity is free. We contribute:
\begin{itemize}
\item \textbf{Interchangeable pods through late binding (\S\ref{sec:latebinding}).} Pods are built from a shared base image per hardware pool (SKU); each job's code, incremental dependencies, and multi-pod topology are bound when it starts, so many distinct experiments share a few warm pools without fragmenting idle capacity.
\item \textbf{Warmth that is cheap and safe (\S\ref{sec:reuse}--\S\ref{sec:pricing}).} \fp{} reuses a pod only for the caller that used it last and retires it after the first job of a new session; a zero-VRAM standby pod waits on a busy GPU in a fractional scheduling share to take over between jobs; and per-accelerator budgets with queue-driven scaling cap how many GPUs warm pools may hold. We also describe the alternatives we rejected.
\item \textbf{Evaluation (\S\ref{sec:eval}).} On production H200 nodes with fixed-duration jobs, agent and RL loops finish 10.8--13\x{} faster than with a fresh pod per job and standby hands a GPU across a reuse boundary in under a second; locally and in simulation, a keep-alive matched to the loop's pauses holds about half the capacity of an always-on pool. Fault and scale tests cover the control plane.
\item \textbf{Production experience (\S\ref{sec:production}).} Across our production clusters, \fp{} served 94.7\% of pod requests in a sampled peak period from warm or standby capacity, and under bursts jobs waited for quota rather than for pods. App.~D gives operating guidance.
\end{itemize}

\section{Agentic and RL Loops and Their Sandboxes}
\label{sec:background}

\subsection{The Workload}

We target two loop shapes. An \textbf{RL loop} fans out parallel rollout jobs, then evaluation jobs, then a policy update, with a barrier between stages, so the slowest pod start in each stage delays the whole iteration. An \textbf{agent loop} alternates short tool executions in isolated sandboxes with LLM think-time pauses. In both, jobs last from seconds to hours or days, need an isolated environment with known GPUs and mounts, and arrive repeatedly from the same caller.

The caller is a program, which shapes the API. A program tolerates queueing as long as its job starts promptly once a GPU is free, but it retries blindly on ambiguous errors. \fp{} therefore rejects unknown request fields, separates a job (the intent) from its attempts, and reports every terminal failure with an error class and exit code (App.~A.6).

\subsection{Where a Cold Start Goes}

\begin{table}[t]
\caption{Anatomy of a cold start on a production H200 node with the 12.2\,GB image cached on the node (88 pods). Phase percentiles do not sum to the total.}
\label{tab:coldstart}
\centering\small
\begin{tabular}{@{}lrr@{}}
\toprule
Phase (pod creation $\rightarrow$ first instruction) & Median & p90 \\
\midrule
Scheduling and GPU device binding & 3.8\,s & 12.8\,s \\
Platform init containers & 14.0\,s & 17.0\,s \\
\quad of which credential fetch & 5.0\,s & 8.5\,s \\
Mounts, image check (around init) & 2.0\,s & 2.0\,s \\
In-pod agent ready, job started & 2.0\,s & 3.1\,s \\
\midrule
\textbf{Total, image cached} & \textbf{22.1\,s} & \textbf{34.7\,s} \\
Create$\to$Ready, 8 at once (1 pod: 19.8\,s) & 47.5\,s & -- \\
\midrule
Fleet mean with image pull, 1-GPU pods & \multicolumn{2}{r}{1.3--3.4\,min} \\
Fleet mean with image pull, 8-GPU pods & \multicolumn{2}{r}{1.7--7.5\,min} \\
\bottomrule
\end{tabular}
\end{table}

With the image cached, 20--35\,s remain (Table~\ref{tab:coldstart}). The largest share is the platform init containers that every production pod in our clusters runs, including a 5\,s credential fetch; a framework's own CUDA initialization comes on top. When eight pods start together on one node, we observed node-level init contention more than double create-to-Ready time, from about 20\,s to nearly 50\,s, with the init chain stalling after the credential fetch. Without the image, the network pull dominates: over a month on our production clusters, we observed average pod-creation times of 1.3--3.4 minutes for 1-GPU shapes and 1.7--7.5 minutes for 8-GPU shapes, with p90 up to 10 minutes. Faster image distribution~\cite{wang2021faasnet}, microVMs~\cite{agache2020firecracker}, checkpoint loaders~\cite{fu2024serverlessllm,zhang2025blitzscale} and GPU checkpoint/restore~\cite{nvidia2024cudacheckpoint,phoenixos} each shorten one step. Reusing an initialized pod removes all of these steps from the critical path.

Caching every custom image on nodes does not scale: each team, framework and pin set yields a distinct 10--30\,GB image, and a node's disk holds only a handful.

\subsection{Three Goals in Tension}

\begin{table}[t]
\caption{How common approaches handle bursts of short GPU jobs. Characterizations of other systems reflect their public documentation. \fp{}'s pool policy reproduces the first three regimes for controlled comparisons (\S\ref{sec:pricing}).}
\label{tab:approaches}
\centering\small
\setlength{\tabcolsep}{3pt}
\begin{tabular}{@{}p{2.85cm}p{1.45cm}p{1.35cm}p{1.6cm}@{}}
\toprule
Approach & Fast start & Few idle GPUs & State isolated between jobs \\
\midrule
New pod per job (K8s Job) & No & Yes & Yes \\
Persistent workers (Ray, static pool) & Yes & No & No \\
Serverless keep-alive & After first job & Partly & Per function \\
Warm-pool controllers (agent-sandbox, SageMaker) & For matching jobs & Per pool & Per claim \\
Images pre-cached on nodes & Removes pull only & Yes & Yes \\
\textbf{\fp{}} & \textbf{Yes, when warm} & \textbf{Capped by budget} & \textbf{Per caller} \\
\bottomrule
\end{tabular}
\end{table}

Fast starts, few idle GPUs and isolation between jobs pull against each other (Table~\ref{tab:approaches}): warm GPUs cost GPU-hours, reuse can carry state between jobs, and every boundary that retires a pod costs a cold replacement unless something hides it. These trade-offs matter only if a warm pod can serve the next job at all (\S\ref{sec:latebinding}).

\section{Design Overview}
\label{sec:overview}

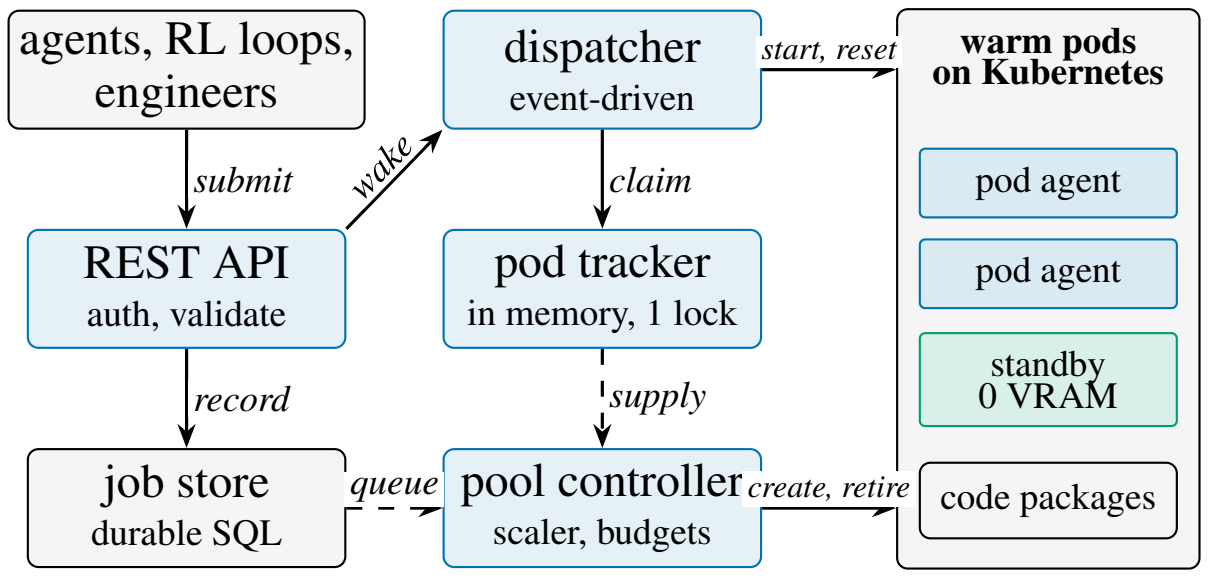
\begin{figure}[t]
\centering
\begin{tikzpicture}[
  font=\footnotesize,
  box/.style={draw, rounded corners=2pt, align=center, minimum width=2.1cm, minimum height=0.78cm, inner sep=2pt},
  core/.style={box, fill=fpblue!10, draw=fpblue},
  ext/.style={box, fill=black!4},
  sub/.style={font=\scriptsize\color{black!70}},
  arr/.style={-{Stealth[length=1.6mm]}, semithick},
  lbl/.style={font=\scriptsize\itshape, inner sep=1pt, fill=white}]
\node[ext] (callers) at (0,0) {agents, RL loops,\\engineers};
\node[core] (api) at (0,-1.45) {REST API\\[-1pt]{\scriptsize auth, validate}};
\node[ext] (store) at (0,-2.9) {job store\\[-1pt]{\scriptsize durable SQL}};
\node[core] (disp) at (2.75,0) {dispatcher\\[-1pt]{\scriptsize event-driven}};
\node[core] (tracker) at (2.75,-1.45) {pod tracker\\[-1pt]{\scriptsize in memory, 1 lock}};
\node[core] (scaler) at (2.75,-2.9) {pool controller\\[-1pt]{\scriptsize scaler, budgets}};
\node[ext, minimum width=2.0cm, minimum height=3.7cm] (pods) at (5.7,-1.45) {};
\node[font=\scriptsize\bfseries, anchor=north, align=center] at (pods.north) {warm pods\\[-2pt]on Kubernetes};
\node[draw=fpblue, fill=fpblue!15, rounded corners=1pt, minimum width=1.7cm, font=\scriptsize] (p1) at (5.7,-0.75) {pod agent};
\node[draw=fpblue, fill=fpblue!15, rounded corners=1pt, minimum width=1.7cm, font=\scriptsize] (p2) at (5.7,-1.35) {pod agent};
\node[draw=fpgreen, fill=fpgreen!15, rounded corners=1pt, minimum width=1.7cm, font=\scriptsize, align=center] (p3) at (5.7,-2.05) {standby\\[-2pt]0 VRAM};
\node[ext, minimum width=1.7cm, minimum height=0.5cm, font=\scriptsize] (pkg) at (5.7,-2.85) {code packages};
\draw[arr] (callers) -- node[lbl, right]{submit} (api);
\draw[arr] (api) -- node[lbl, right]{record} (store);
\draw[arr] (api.north east) -- node[lbl, sloped, above]{wake} (disp.south west);
\draw[arr] (disp) -- node[lbl, right]{claim} (tracker);
\draw[arr, dashed] (tracker) -- node[lbl, right]{supply} (scaler);
\draw[arr, dashed] (store) -- node[lbl, above]{queue} (scaler);
\draw[arr] (disp.east) -- node[lbl, above, font=\tiny\itshape]{start, reset} (pods.west |- disp.east);
\draw[arr] (scaler.east) -- node[lbl, above, font=\tiny\itshape]{create, retire} (pods.west |- scaler.east);
\end{tikzpicture}
\caption{\fp{} architecture. Callers submit jobs to one broker per Kubernetes cluster, which records them in a durable job store. The dispatcher claims a warm pod from an in-memory tracker in one atomic step and starts the job through the pod's agent, which fetches the job's code package. A pool controller creates and retires pods from queue depth and budgets.}
\label{fig:arch}
\end{figure}

\fp{} runs as one control-plane service per Kubernetes cluster (Figure~\ref{fig:arch}). A caller's job names a hardware pool (SKU) and, optionally, a code package; the SKU fixes the GPU model, CPU, memory, base image and scaling policy. The broker records the job in a durable SQL store (MySQL in production). A dispatch loop takes queued jobs in priority order, claims warm pods from an in-memory tracker, and starts each job through an executor that talks to a small agent in the pod. Submission, pod release and pod readiness wake the loop; a 3\,s tick remains as a fallback. A pool controller creates and retires pods to match queued demand.

\para{One lock, held for microseconds.} The tracker indexes every pod by name, SKU, caller and job behind one read/write lock. Finding an eligible pod, validating the transition, checking per-caller limits and recording the lease happen in one step under the write lock, so that, by design, no two workers claim the same pod or overrun a quota. On a quiet host the lock is held 1.5--2.6\,$\mu$s at the median ($\leq$8\,$\mu$s p99) with 100{,}000 tracked pods, sustaining 110k--220k claims/s, and the index costs about 1\,KB per pod (\S\ref{sec:eval-scale}). The durable commit that follows the claim, not the lock, bounds dispatch. No user code runs before this version-checked commit, and a failed commit returns the pods. A restarted broker rebuilds the tracker before dispatching, trusting the job store for jobs and pod annotations for ownership (App.~A.1).

\rejected{one lock per SKU.} Several invariants span SKUs (per-caller GPU limits, per-accelerator budgets, standby exclusivity), so per-SKU locks would need a multi-lock order; one lock held for 2\,$\mu$s is simpler and far cheaper than the commit that follows.

\section{Interchangeable Sandboxes: Bind the Job Late}
\label{sec:latebinding}

A warm pool helps only if arriving jobs can use its pods. If every image, requirement set or topology needed its own pool, warm capacity would splinter and most starts would be cold. \fp{} keeps pods interchangeable within each SKU by binding as little as possible when a pod is created and as much as possible when a job is assigned.

\para{A shared base image per pool.} Each SKU pool uses one curated image with the heavy, slow-changing layers, so the cluster caches a few base images per GPU generation instead of every experiment's.

\para{Code and dependencies arrive with the job.} User code travels as a content-addressed zip package, uploaded once and unpacked into the pod's workspace at job start (about 0.4\,s at the median for packages of a few megabytes). A bootstrap step installs extra Python dependencies on the caller's first job on a pod and keeps them for that caller's later jobs, which owner-only reuse (\S\ref{sec:reuse}) sends back to the same pod. The install is keyed by a digest of the requirements file, and a pod remembers only the last set installed.

\para{Topology arrives with the job.} Kubernetes training operators usually bake ranks into pod specs; \fp{} pods boot as generic workers. When a job asks for $N$ pods, the dispatcher claims all $N$ in one step under the tracker lock, so no job holds a partial set that could deadlock against another. Without reservations, a large gang (a multi-pod job) could starve behind sustained small jobs (App.~A.2). The executor then designates a coordinator, injects the \code{torchrun} rendezvous and rank variables (or starts Ray head and worker daemons), and watches for completion. When the job ends, the pods' reset strips the ephemeral cluster state and they return to the pool as independent warm workers.

\para{Where the boundary falls.} Native builds and framework pins belong in the image, where incompatibilities surface once at build time; prompts, rewards, verifiers and experiment logic belong in the package, where an edit takes effect on the next job.

\para{Costs and limits.} Packages are unpacked on every job, so they are capped (512\,MiB compressed, 2\,GB extracted) and should stay small relative to job time (\S\ref{sec:eval-latebinding}). Incremental dependencies are added or upgraded in place and never uninstalled until the pod retires; a caller who needs a clean environment asks for a single-use pod (App.~A.2). Callers who must bring their own image may do so from an allowlist: a mutable tag runs on a single-use pod, and a digest-pinned image may be reused by the same caller but never joins the shared SKU pool.

\rejected{exact shapes or per-image pools.} Arbitrary shapes, or a separate pool for each image, fragment warm capacity and turn constant-time bucket lookups into best-fit searches under the lock; \fp{} rounds requests up to the smallest fitting SKU (\S\ref{sec:eval-latebinding}).

\section{Safe Reuse and Isolation Boundaries}
\label{sec:reuse}

Reusing a pod blindly carries files, shared memory or credentials from one job to the next. \fp{} bounds what crosses with four rules, decided at \emph{claim} time so the release path stays fast.

\para{1. Owner-only reuse and owner holds.} In production, a used pod is reused only for the caller that used it last. To place a job, the broker searches the SKU bucket in order: an idle pod last used by the \emph{same caller} (keeping its page cache and installed dependencies); a never-used idle pod; a zero-VRAM standby on an active GPU (\S\ref{sec:standby}); and only then a new pod. When a job finishes, the pod stays reserved for its caller for an \emph{owner hold} (\code{caller\_hold\_seconds}, configurable per pool), so the caller's next loop step reclaims it before anyone else can.

\para{2. Retire at the session boundary.} A caller marks the steps of one session with a reuse tag (\code{lease\_tag}), and each pod records the tag and queue of its last job. Jobs from the same caller, queue and tag reuse the warm pod. When the caller starts a new session (a new tag) or switches queue, its next job may run once on the old pod so it does not wait, and the broker retires that pod when the job finishes, without consulting the database. A pod whose owner hold expires is also retired. State from one session can therefore reach at most one job of the same caller's next session; by default, pods are never reused across callers. Operators can instead configure the tag as a soft routing hint that prefers the same pod without retiring it (\S\ref{sec:eval-isolation}); caller and queue boundaries still apply.

\para{3. Verified reset between jobs.} Between jobs, the in-pod agent kills the job's entire process tree and verifies against \code{/proc} that it is gone; the exiting processes close every file descriptor and release the CUDA context and GPU memory. It then empties the package directory, \code{/dev/shm}, per-job GPU accounting files and any volume marked for wiping, while keeping read-only caches. If the agent cannot confirm a clean reset, for example a process that survives \code{SIGKILL}, the broker deletes the pod instead of reusing it.

\para{4. Verified caller identity.} Callers authenticate with cluster-issued or federated identity tokens (validation takes 6.6\,$\mu$s when cached). The broker injects the verified identity into the job and filters user-supplied environment variables that could affect the agent or the job's loader.

\section{Standby: Handing Over a GPU Without a New Pod}
\label{sec:standby}

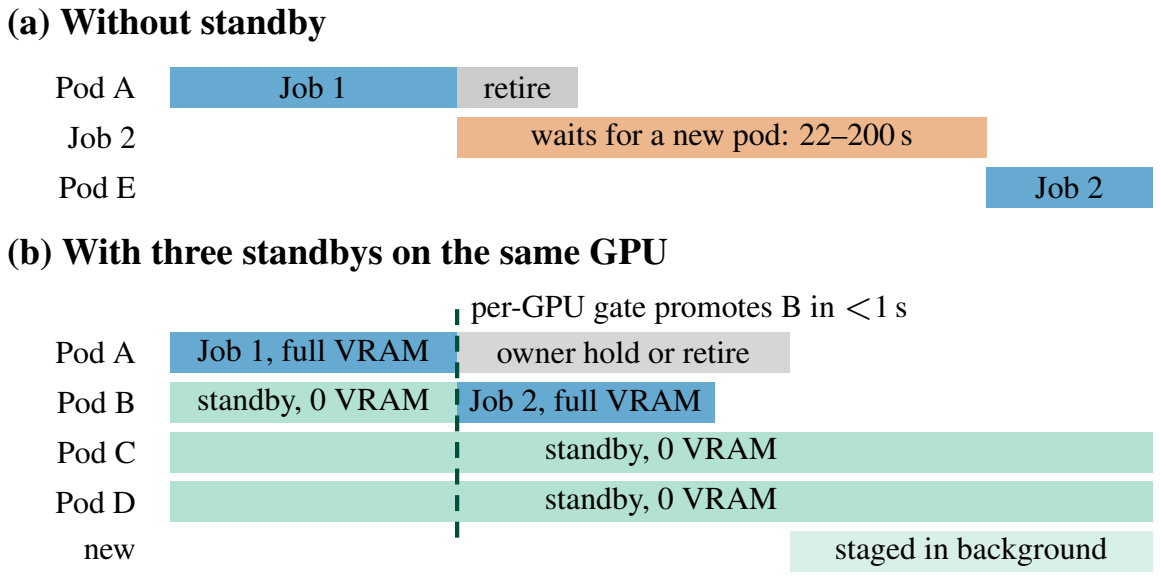
\begin{figure}[t]
\centering
\begin{tikzpicture}[font=\scriptsize, x=1cm, y=1cm]
\def\h{0.27}
\providecommand{\lanebar}[5]{\fill[#4] (#2,{-#1*0.33}) rectangle (#3,{-#1*0.33+\h}); \node[font=\tiny] at ({(#2+#3)/2},{-#1*0.33+\h/2}) {#5};}
\node[anchor=west, font=\scriptsize\bfseries] at (-0.1,0.55) {(a) Without standby};
\foreach \l/\t in {0/Pod A, 1/Job 2, 2/Pod E} \node[anchor=east, font=\tiny] at (1.0,{-\l*0.33+\h/2}) {\t};
\lanebar{0}{1.1}{3.0}{fpblue!60}{Job 1}
\lanebar{0}{3.0}{3.8}{fpgrey!50}{retire}
\lanebar{1}{3.0}{6.5}{fpverm!45}{waits for a new pod: 22--200\,s}
\lanebar{2}{6.5}{7.6}{fpblue!60}{Job 2}
\begin{scope}[yshift=-1.75cm]
\node[anchor=west, font=\scriptsize\bfseries] at (-0.1,0.78) {(b) With three standbys on the same GPU};
\foreach \l/\t in {0/Pod A, 1/Pod B, 2/Pod C, 3/Pod D, 4/new} \node[anchor=east, font=\tiny] at (1.0,{-\l*0.33+\h/2}) {\t};
\lanebar{0}{1.1}{3.0}{fpblue!60}{Job 1, full VRAM}
\lanebar{0}{3.0}{5.2}{fpgrey!40}{owner hold or retire}
\lanebar{1}{1.1}{3.0}{fpgreen!30}{standby, 0 VRAM}
\lanebar{1}{3.0}{4.7}{fpblue!60}{Job 2, full VRAM}
\lanebar{2}{1.1}{7.6}{fpgreen!30}{standby, 0 VRAM}
\lanebar{3}{1.1}{7.6}{fpgreen!30}{standby, 0 VRAM}
\lanebar{4}{5.2}{7.6}{fpgreen!15}{staged in background}
\draw[densely dashed, fpgreen!50!black, thick] (3.0,0.42) -- (3.0,-1.12);
\node[anchor=west, font=\tiny, fill=white, inner sep=1pt] at (3.05,0.42) {per-GPU gate promotes B in $<$1\,s};
\end{scope}
\end{tikzpicture}
\caption{Closing the gap between boundary-crossing jobs on one GPU. (a) When Pod A must be held for its owner or retired, Job 2 waits for a new pod. (b) Standby pods B--D share A's GPU in fractional scheduler slots but run no CUDA context and hold no VRAM. When Job 1's processes are gone, a per-GPU gate promotes B with the whole card's memory while A is held or retired off the critical path.}
\label{fig:standby}
\end{figure}

When the next job comes from a \emph{different} caller, or crosses a reuse-tag boundary, the previous pod must be held or retired, and on a capacity-capped pool the job waits 22--200\,s for a replacement (Figure~\ref{fig:standby}a).

\para{Zero-VRAM standby.} GPU sharing usually lets several jobs \emph{run} at once~\cite{yu2020salus,xiao2020antman}; \fp{} uses fractional scheduling instead to overlap the next pod's provisioning with the current job (Figure~\ref{fig:standby}b). On our H200 clusters the GPU-sharing scheduler (HAMi~\cite{hami2024} with Volcano~\cite{volcano2024}) allocates fractional shares of a card. For SKUs with standby enabled, \fp{} places one active pod and up to three standby pods on the same physical GPU (\code{standbys\_per\_active} is configurable), pinned to the active's card by GPU UUID. The share is only a scheduler slot: a standby runs nothing but the idle pod agent, creates no CUDA context and allocates no GPU memory. When a job runs on the active pod, the agent raises the CUDA interception library's per-GPU limit to the whole card in the job \emph{process}'s environment, never in the pod spec, so the job can use all 141\,GB of the H200 while standbys keep their fractional cap.

\para{Atomic promotion behind a per-GPU gate.} A standby is promoted only for a dispatched job, which claims it under the tracker lock when no same-owner or unheld idle pod is available. A per-GPU gate lets at most one pod per physical GPU run a job at a time, so the gate promotes the claimed standby once the previous job has exited. A pod reset for reuse is released only after its process tree is verified dead (\S\ref{sec:reuse}), and the driver frees a context's memory when its processes exit. The controller later stages a fresh standby in the open slot.

\para{Counting the physical GPU.} \fp{} counts each active--standby group as one physical GPU in every budget; counting its up to four pods separately would let held pods spill standbys onto extra GPUs. Before promoting, the broker checks that every share on the card belongs to that \fp{} group; after a restart, promotions fail closed until a live sweep re-verifies exclusivity (App.~A.4).

\section{Pricing Warmth: Queue-Driven Scaling and Budgets}
\label{sec:pricing}

Warm pods cost GPU time, so every pool has explicit budgets counted in physical GPUs (or CPU cores), and 1-GPU and 8-GPU pools share one per-accelerator ledger.

\para{Scale on the queue, not on utilization.}%
 Warm pods are idle by design, so a utilization-driven scaler would shrink the pool just when it is ready to absorb a burst. \fp{} instead scales each pool from queued demand and current supply, in whole pods, on every reconcile pass (30\,s by default). It creates $\mathit{add}$ pods, where
{\small\begin{align*}
\mathit{add} &= \min(\mathit{need},\ \mathit{ceiling}-\mathit{physical},\ \mathit{step},\ \mathit{headroom}),\\
\mathit{need} &= \max(\mathit{floor}-\mathit{physical},\ \mathit{demand}-\mathit{eligible},\ 0),
\end{align*}}%
and \emph{floor} and \emph{ceiling} are pool knobs defined below. \emph{Demand} counts queued, assigned and running pods; \emph{eligible} counts starting, idle, assigned and running pods; and \emph{physical} adds owner-held and stale pods to the eligible ones (App.~A.5).
All terms count GPUs rounded to whole pods, and an active--standby group counts once. \emph{Queued} includes jobs waiting on their caller's quota; the ceiling bounds the overshoot this causes. \emph{Stale} pods (old templates, single-use pods) occupy capacity but cannot serve warm jobs. \emph{Step} caps pod creations per pass, and \emph{headroom} caps them by the global budget. Two accounting choices prevent oscillation and starvation. \emph{Starting pods are supply}: GPU pods take 22--200\,s to become ready, and without this the controller would create another wave on every pass. \emph{Owner-held pods count toward the floor and the ceiling but are not supply for other callers}: a held pod is reserved for its previous caller, so counting it as supply would starve the rest of the queue, while leaving it out of the physical count would overshoot the floor and the ceiling. Held pods that match their owner's own queued jobs count as idle. When a queued job is blocked at the ceiling, the broker breaks expired or excess holds on pods older than a minimum tenure (60\,s) and retires those pods, sparing owners with queued jobs. Scale-down removes only idle pods whose keep-alive has expired, never below the floor, and retains up to the idle cap (App.~A.5).

\para{Four knobs.} Operators tune each pool's latency--cost trade-off with a \emph{floor} (\code{min\_count}, pods kept ready at zero demand), an \emph{idle cap} (\code{max\_idle\_count}, unheld idle pods kept past their keep-alive; it retains but never pre-creates), a \emph{ceiling} (\code{max\_count}, a hard cap on GPUs held), and a \emph{keep-alive} $\tau$ (\code{idle\_timeout}, with the separate owner hold \code{caller\_hold\_seconds}). Floor $=$ idle cap $=0$ with keep-alive $\tau$ is serverless TTL retention; floor $=$ peak with $\tau=\infty$ is a static cluster; single-use pods are Kubernetes Jobs (\S\ref{sec:eval-cost}).

\para{Hard limits and borrowed capacity.} A deployment caps each GPU model across all pools, reserving GPUs in an in-memory ledger \emph{before} calling the Kubernetes API, so concurrent reconcilers never overshoot. \fp{} can also place warm or burst pods on unused reserved capacity, as advertised by the cluster's capacity controller, and gives them back within about 30\,s when the owner reclaims its quota (\S\ref{sec:eval-scale}, App.~A.7).

\section{Production Experience}
\label{sec:production}

\fp{} runs at LinkedIn across our production Kubernetes clusters, one broker per cluster, managing CPU pools and multiple generations of NVIDIA data-center GPUs. It powers coding-agent, RL training (fanning out 100+ parallel rollout jobs per step) and other agentic research workloads.

\begin{table}[t]
\caption{Warm-or-standby service in production over a sampled 14-hour peak period, including a 3-hour burst of $\sim$2,500 single-GPU H200 jobs. Pool rows are broker counter rates; the standby row counts per-job records.}
\label{tab:prodhit}
\centering\small
\begin{tabular}{@{}lr@{}}
\toprule
Scope & Warm or standby \\
\midrule
All pools (3,206 acquisitions) & \textbf{94.7\%} \\
Single-GPU H200 pool & \textbf{98.6\%} \\
Second single-GPU H200 pool & 97.7\% \\
Standby when idle pods ran out & 187/222 (84.2\%) \\
\bottomrule
\end{tabular}
\end{table}

\para{94.7\% of pod requests are served warm.} Over a sampled 14-hour peak period, \fp{} served 94.7\% of 3,206 pod acquisitions (98.6\% on the single-GPU H200 pool) without creating a pod (Table~\ref{tab:prodhit}). When a 3-hour burst exhausted that pool's unheld idle pods, standby promotions served 187 of the 222 jobs that found none (84.2\%); the other 35 got new pods. After the burst, a floor of two idle pods absorbed steady arrivals.

\begin{table}[t]
\caption{Start latency by how the pod was obtained, single-GPU H200 pools, one week. \emph{Submit}$\to$\emph{first} is what the caller sees, including queueing for quota; \emph{pod}$\to$\emph{first} is setup after a pod is assigned. Quiet: an uncontended pool; burst: the burst above, where new base-image pods are not broken out.}
\label{tab:prodpath}
\centering\small
\setlength{\tabcolsep}{3.5pt}
\begin{tabular}{@{}llrr@{}}
\toprule
Load & Pod obtained by & Submit$\to$first & Pod$\to$first \\
 & & p50 / p90 (s) & p50 / p90 (s) \\
\midrule
Quiet & Warm reuse & 1.8 / 3.5 & 0.5 / 0.7 \\
 & Standby promotion & 2.4 / 50.4 & 0.5 / 0.7 \\
 & New pod (base image) & 65.2 / 114.4 & 0.6 / 0.7 \\
Burst & Warm reuse & 53.0 / 267.6 & 0.4 / 0.6 \\
 & Standby promotion & 112.1 / 278.6 & 0.3 / 0.5 \\
 & New pod (custom image) & 69.7 / 234.3 & 54.1 / 171.6 \\
\bottomrule
\end{tabular}
\end{table}

\para{Under a burst, jobs wait for quota, not for pods.} Once a job is assigned a warm or standby pod, its code starts in 0.3--0.5\,s at the median and $\leq$0.7\,s at p90, quiet or burst (Table~\ref{tab:prodpath}). During the burst, warm-reuse jobs waited 53\,s at the median because callers had saturated their GPU quota and queued behind their own jobs; budgets and borrowing (\S\ref{sec:pricing}) govern that wait. Only custom-image jobs wait after assignment (54.1\,s), while their image is pulled; their lower burst wait reflects different callers and queues, not faster pods. The quiet standby p90 reflects retry polling from before dispatch became event-driven.

\para{What warmth costs in production.} Idle pods account for a small share of the GPU time \fp{} holds in production. Most of that is floors kept on low-traffic pools, which operators can lower; owner holds add little. The broker adds only a few requests per second to each cluster's API server and did not trigger client throttling in our observations.

\section{Evaluation}
\label{sec:eval}

\para{Venues.} Headline results come from controlled benchmarks in dedicated pools on production nodes: \textbf{H200} uses one pinned 8\x{}H200 node with 1-GPU pods, a cached 12.2\,GB base image and a two-GPU pool cap, and \textbf{CPU} uses 4\,vCPU/8\,GiB pods with a cached 4.4\,GiB image and a 16-pod cap. Mechanism costs come from a local cluster, and large parameter sweeps from a trace-calibrated simulator of \fp{}'s control plane (App.~B).

\para{Workloads and arms.} \textbf{W-RL} runs 4 barrier-synchronized iterations of 8 rollouts, 2 evaluations and 1 policy update (fixed 1.5--4\,s bodies). \textbf{W-Agent} runs 2 or 4 concurrent agents for 3 iterations of 2--4 calls each (fixed 0.5--1.6\,s bodies, the last a verify step), with 0.5--2\,s think pauses between calls and 2--5\,s between iterations. We compare three arms. \textbf{Reuse On} gives each caller a stable reuse tag. \textbf{Reuse Off} gives every job a fresh tag, so a pod runs at most one job past its own tag and then retires, and a job may also land on an existing idle, recycled or standby pod; it ablates owner reuse. \textbf{Fresh pod per job} makes \fp{} create a single-use pod for each job and delete it at release, the closest analogue of a Kubernetes Job. Arms are interleaved in a Williams crossover with 6--10 paired repeats; we report pooled percentiles and paired differences with 95\% bootstrap CIs.

\subsection{Is Pod Creation Off the Critical Path?}
\label{sec:eval-critical}

\begin{table*}[t]
\caption{Warm reuse against a fresh pod per job (H200) and against per-job reuse tags (H200 and CPU), production nodes. Percentiles pool all jobs; the paired difference is the median of per-repeat p50s, Off/Fresh minus On, with 95\% bootstrap CI. Makespan is the median over repeats. (N) counts jobs over both arms. Agent columns ran 4 agents, except H200 vs Reuse Off (2). The fresh-pod columns ran the event-driven dispatcher of \S\ref{sec:overview}; the Reuse Off columns ran an earlier, tick-driven one with the reuse tag as a retiring boundary (\S\ref{sec:reuse}). Speedups use unrounded medians.}
\label{tab:critical}
\centering\footnotesize
\setlength{\tabcolsep}{3pt}
\begin{tabular}{@{}lccccc@{}}
\toprule
 & \multicolumn{2}{c}{H200 vs fresh pod per job} & \multicolumn{3}{c}{vs Reuse Off (fresh tag per job)} \\
\cmidrule(lr){2-3}\cmidrule(l){4-6}
 & W-Agent (720) & W-RL (880) & H200 Agent (380) & CPU RL (528) & CPU Agent (444) \\
\midrule
Start p50 (s), On / Off & \textbf{0.61 / 89.5} & \textbf{3.92 / 115.8} & 1.91 / 2.29 & \textbf{2.83 / 52.6} & \textbf{1.78 / 27.5} \\
Start p90 (s), On / Off & \textbf{3.51 / 354.6} & \textbf{80.8 / 315.2} & \textbf{3.52 / 59.1} & \textbf{29.9 / 62.0} & 52.8 / 58.0 \\
Paired p50 diff.\ (s) & +95.4 [82.4, 107.8] & +119.1 [105.5, 134.4] & +0.43 [$-$0.03, 0.84] & +43.7 [36.2, 49.1] & +25.5 [24.9, 26.1] \\
Makespan (s), On / Off & \textbf{119.3 / 1,554.2} & \textbf{163.0 / 1,761.9} & \textbf{51.8 / 186.4} & \textbf{116 / 516} & \textbf{186 / 403} \\
Loop speedup & \textbf{13.0\x} & \textbf{10.8\x} & \textbf{3.6\x} & \textbf{4.5\x} & \textbf{2.2\x} \\
\bottomrule
\end{tabular}
\end{table*}

\begin{figure*}[t]
\centering
\includegraphics[width=\textwidth]{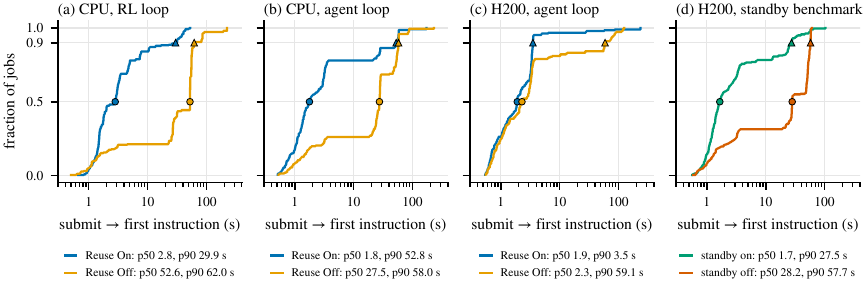}
\caption{Start-latency distributions, all repeats pooled (circles p50, triangles p90). (a, b) On CPU, Reuse Off shifts most jobs right by a pod start. (c) On H200 the medians are close because Reuse Off jobs often land on standby or recycled pods; the gap is the tail of jobs that wait for a new pod under the two-GPU cap. (d) The standby benchmark (\S\ref{sec:eval-standby}), a fresh reuse tag per job, all jobs.}
\label{fig:cdf}
\end{figure*}

\textbf{Yes. With fixed-duration jobs on H200 and against a fresh pod per job, warm reuse finishes agent and RL loops 13.0\x{} and 10.8\x{} faster; against per-job reuse tags, it removes the tail (p90 59.1\,s $\rightarrow$ 3.5\,s).}

\para{Fresh pod per job.} Under a two-GPU cap, a single-use pod per job admits two jobs at a time and each waits for the previous pod's deletion and a new pod, so a job queues about 90\,s at the median and an end-to-end multi-step agent loop takes 22--32 minutes instead of 0.6--2.5 minutes (Table~\ref{tab:critical}; paired differences in App.~C.6). Every one of the 1,600 jobs succeeded, with zero image pulls. The warm RL p90 (80.8\,s) comes from rollouts queueing behind the two busy GPUs.

\para{Reuse Off.} This arm removes only owner reuse. The H200 pool keeps three zero-VRAM standby pods per active GPU, so standby or recycled pods served 137 of the arm's 190 jobs, and its median start latency is close to Reuse On's (the paired CI spans zero). The 20 jobs that waited for a new pod (82\,s median), and those queued behind them, make the loop 3.6\x{} slower (paired +130\,s [74, 141]). On CPU, which has no standby, Reuse Off pays a pod start on most jobs (+43.7\,s and +25.5\,s at p50), and W-RL's barriers turn these stragglers into a 4.5\x{} slower loop (Figure~\ref{fig:cdf}a).

\subsection{Does Late Binding Keep Pods Interchangeable?}
\label{sec:eval-latebinding}

\begin{figure*}[t]
\centering
\includegraphics[width=\textwidth]{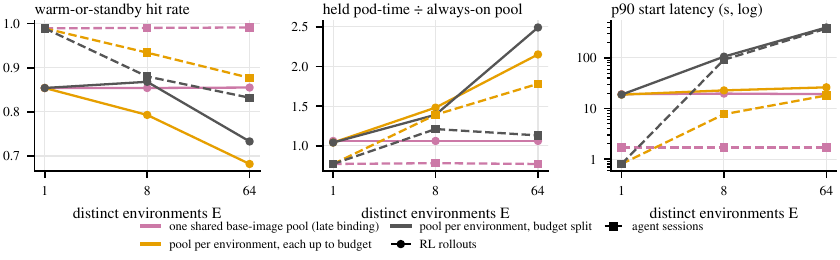}
\caption{Environment fragmentation (simulator, \fp{}'s control plane, CPU calibration with a 22\,s pod start, 64-pod budget, Zipf-distributed environments, 10 paired seeds). One late-bound pool stays flat as the number of distinct environments $E$ grows; a pool per environment roughly doubles held capacity when each may grow to the budget, and with the budget split it sharply raises tail latency while still holding more than the shared pool.}
\label{fig:lb6}
\end{figure*}

\textbf{Yes: binding a job costs about 0.4\,s, and one shared pool stays flat as environments multiply.}

Across the production benchmarks, delivering the code package, unpacking it and starting the job took 0.40--0.54\,s at the median (p99 $\leq$1.45\,s on H200).

Late binding's larger benefit is the fragmentation it avoids (Figure~\ref{fig:lb6}). With 64 environments, a pool per environment that may grow to the budget sees its RL-rollout hit rate fall from 0.85 to 0.68 and holds 2.15\x{} an always-on pool, against 1.06\x{} for one shared pool; splitting the budget instead pushes p90 start latency to 400\,s. The shared pool loses only if binding is slow: with a 30\,s hand-off it holds as much as per-environment pools.

\subsection{Is the Gap Between Boundary-Crossing Jobs Gone?}
\label{sec:eval-standby}

\begin{table}[t]
\caption{Standby on H200: 2-GPU pool, 10 paired repeats of 30 sequential jobs with a fresh reuse tag each (300 jobs per arm). A \emph{handover} is a job that found no idle pod. Queue wait is over handovers; submit$\to$first pools all jobs. The makespan difference is the median of paired differences.}
\label{tab:standby}
\centering\small
\setlength{\tabcolsep}{3pt}
\begin{tabular}{@{}lrr@{}}
\toprule
 & Standby on & Standby off \\
\midrule
Handovers served by standby & 101/117 (86\%) & 0/156 \\
Queue wait at handover, median & \textbf{0.94\,s} & 51.0\,s \\
Submit$\to$first instr., p50 & 1.66\,s & 28.2\,s \\
Loop makespan, median (3.0\x) & \textbf{337\,s} & 1,023\,s \\
\bottomrule
\end{tabular}
\end{table}

\textbf{Yes: standby promotion cuts the handover wait from 51\,s to 0.94\,s and the loop 3.0\x.} One caller runs 30 sequential agent jobs, each with a fresh reuse tag (Table~\ref{tab:standby}). A job may reuse the previous job's pod once, after which that pod is retired, so jobs alternate between reusing the previous pod and a handover. With standby, 86\% of handovers were promotions; the median queue wait over handovers was 0.94\,s (0.81\,s for promotions alone), after which hand-off and first instruction add 0.5\,s. All 10 paired repeats favored standby (paired makespan +685\,s [591, 766]). Figure~\ref{fig:cdf}d shows the distributions.

\subsection{What Does Warmth Cost?}
\label{sec:eval-cost}

\begin{figure*}[t]
\centering
\includegraphics[width=\textwidth]{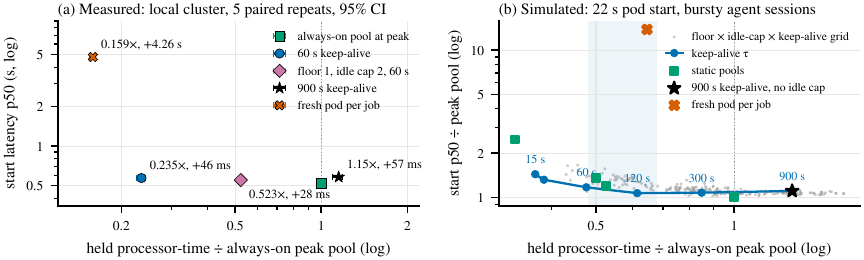}
\caption{Held capacity against start latency, relative to an always-on pool sized to peak concurrency (lower-left is better). (a) Measured on the local cluster (2,300 jobs). (b) Simulated (CPU calibration) with the production 22\,s pod start; the shaded band is a 60--120\,s keep-alive. A 900\,s keep-alive holds more than an always-on pool in both.}
\label{fig:frontier}
\end{figure*}

\textbf{Locally and in simulation, a keep-alive matched to the loop's pauses keeps the latency benefit while holding about half the capacity of an always-on pool; measured on H200, warm reuse costs 15\% more GPU time per job for a 3.6\x{} faster loop.} We measure \emph{held capacity}: GPU- or core-seconds held by a pool's pods from binding to deletion, busy, idle or owner-held, relative to an always-on pool sized to peak concurrency.

On the local cluster (Figure~\ref{fig:frontier}a), 60\,s keep-alives held 0.24--0.52\x{} the always-on pool for +28--46\,ms at the median. With production's 22\,s pod start (Figure~\ref{fig:frontier}b), a 60--120\,s keep-alive holds 0.48--0.68\x{} for +0.01 to +0.27\,s at the median. In contrast, a long 900\,s keep-alive with no idle cap holds 1.15\x{} measured and 1.12--1.54\x{} simulated, as it retains burst overshoot across the entire timeout window. On H200, warm reuse held 12.3 GPU-seconds per job against 10.7 for Reuse Off, mostly pods held for their caller between agent steps.

Warmth stops paying for loops when jobs are long: the per-job saving stays at 10--35\,s on CPU, but the loop speedup falls below 1.1\x{} beyond about 200\,s (RL) or 100\,s (agent) bodies, and vanishes when think time exceeds the keep-alive (App.~C.1).

\subsection{What Does Isolation Cost?}
\label{sec:eval-isolation}

\textbf{The boundary policy dominates; the reset itself takes 9\,ms.} Locally, retiring a pod at each boundary added 1.53\,s [0.97, 2.00] to start latency, and a fresh pod per job added 3.42\,s [2.80, 4.00]. On production GPUs, a retirement costs a 22--200\,s pod creation unless a standby absorbs it. Treating the reuse tag as a retiring boundary kept only about half of multi-step jobs warm; with the tag as a routing hint and the event-driven dispatcher, the warm share rose to 0.99 and pods created per repeat fell from 28.9 to 2.5. The in-pod reset takes 9.3\,ms for a clean job and 1.67\,s in the worst case we built; App.~C.2 also covers injected unrecoverable resets and cross-caller isolation.

\subsection{Does It Hold Under Failure and Scale?}
\label{sec:eval-scale}

\begin{table}[t]
\caption{Control-plane failure and scale tests (local cluster, quiet hosts, kwok).}
\label{tab:scale}
\centering\small
\setlength{\tabcolsep}{3pt}
\begin{tabular}{@{}>{\raggedright\arraybackslash}p{1.7cm}>{\raggedright\arraybackslash}p{6.3cm}@{}}
\toprule
Test & Result \\
\midrule
Broker killed 30\x{} under load & 0 lost, 0 duplicated jobs; 90/90 owner holds restored; dispatching 0.045\,s and first claim 3.0\,s after restart; API reachable 20\,s after the kill \\
Rollout, saturated pool & Converged in 324\,s; 12/12 callers served; 0 failed jobs; cap never exceeded; $\Delta$p50 +0.03\,s vs.\ no rollout \\
Gang claims, $N$=2--8 & 0 partial claims in 50 jobs; warm claim 0.27--0.45\,s \\
Tracker, 100k pods & 110k--220k claims/s; lock 1.5--2.6\,$\mu$s; $\sim$1\,KB/pod \\
Durable dispatch & One replica: MySQL 483, PostgreSQL 2,365 jobs/s \\
kwok, 20k pods & No API throttling; \fp{}'s own create rate binds first \\
Borrowing & Borrow 11.8\,s; return 30.0\,s; owner reclaim 80/80 \\
\bottomrule
\end{tabular}
\end{table}

\textbf{Yes.} Table~\ref{tab:scale} summarizes fault-injection and scale tests validating broker recovery, gang claims and capacity borrowing (detailed in App.~C.3). Durable dispatch bounds throughput, at 483 jobs/s on MySQL, far above the peak arrival rates we have observed on a cluster.

\section{Related Work}
\label{sec:related}

\para{Cold starts and keep-alive.} SOCK~\cite{oakes2018sock}, Catalyzer~\cite{du2020catalyzer}, Firecracker~\cite{agache2020firecracker}, FaaSNet~\cite{wang2021faasnet} and FaaSnap~\cite{ao2022faasnap} cut serverless cold starts with lightweight sandboxes, snapshots and fast image provisioning; ServerlessLLM~\cite{fu2024serverlessllm} and BlitzScale~\cite{zhang2025blitzscale} load model checkpoints fast; Shahrad et al.~\cite{shahrad2020serverless} tune keep-alive windows against memory. GPU checkpoint/restore~\cite{nvidia2024cudacheckpoint,phoenixos} restores an initialized CUDA process without a live pod, but not a container's credentials, mounts and arbitrary process tree. These shorten a cold start; \fp{} removes it from the critical path and composes with them for the starts it cannot avoid. Warm-pool controllers such as Kubernetes \code{agent-sandbox}~\cite{agentsandbox2026}, SageMaker Managed Warm Pools~\cite{aws2026warmpools} and OpenKruise Agents~\cite{openkruiseagents2026} keep pre-warmed sandboxes or instances for \emph{matching} requests (our per-environment baselines in \S\ref{sec:eval-latebinding}), and SpecBox~\cite{zhang2026specbox} pre-warms them speculatively. \fp{} adds late binding, which lets diverse jobs match one pool, as well as zero-VRAM GPU standby, queue-driven GPU budgets, and claims made in memory rather than through the API server.

\para{GPU sharing and switching.} MIG, MPS~\cite{nvidia2024mig,nvidia2024mps}, Salus~\cite{yu2020salus}, AntMan~\cite{xiao2020antman} and HAMi~\cite{hami2024} co-run workloads on a GPU; PipeSwitch~\cite{bai2020pipeswitch} switches contexts between processes in one server. \fp{} uses fractional shares only to stage standbys, handing a GPU across isolated pods.

\para{Cluster schedulers and RL frameworks.} Gandiva~\cite{xiao2018gandiva}, Tiresias~\cite{gu2019tiresias}, Pollux~\cite{qiao2021pollux}, MAST~\cite{choudhury2024mast} and Rubick~\cite{zhang2025rubick} schedule long-running training, where minutes of startup amortize over hours (characterized in~\cite{jeon2019philly,weng2022mlaas}). Kueue~\cite{kueue2024}, Volcano~\cite{volcano2024} and SkyPilot~\cite{yang2023skypilot} manage queueing, quotas and provisioning per job without reusable warm pods. Ray~\cite{moritz2018ray}, HybridFlow~\cite{sheng2025hybridflow}, RLHFuse~\cite{zhong2025rlhfuse} and RollArt~\cite{gao2025rollart} hold or disaggregate workers for a training job; \fp{} brokers the short, isolated tool, verifier and rollout sandboxes around them.

\section{Conclusion}
\label{sec:conclusion}

When agents and RL coordinators replace people in the ML loop, provisioning moves onto the critical path of every iteration. \fp{} takes it off by making pods interchangeable through late binding and keeping them warm safely within GPU budgets. On H200, agent and RL loops finish 10.8--13\x{} faster than with a fresh pod per job.

\para{Limits and next steps.} Warmth does not help jobs that bring their own images, which dominate our 8-GPU pools; image-keyed pools or late binding for multi-GPU images are next. Standby is a buffer that refills at pod-creation speed; extending it to multi-GPU gangs needs gang-wide promotion, and large gangs need reservations to avoid starvation. Reuse scoped to a caller and session, with a verified reset between jobs, is designed to isolate callers from each other, though state from one session can reach the first job of the same caller's next session. \fp{} targets internal workloads; its pools compose with microVMs where kernel-level isolation is required. Recovery is tested against broker crashes, not network partitions, during which a re-queued attempt can overlap its predecessor (App.~A.1).

\section*{AI Usage Disclosure}
The authors designed \fp{} and its experiments. GitHub Copilot (coding agent and CLI, with Claude and GPT models) helped write parts of \fp{}, its benchmark harnesses and its analysis and figure scripts; all code was reviewed by the authors. All reported numbers are computed by those scripts from recorded run artifacts; no data was generated by AI. For the paper, Copilot converted the authors' draft to LaTeX and applied author-directed revisions and wording edits. AI reviewers (DeepReview, \mbox{GPT-6} Astra and Claude Opus 5.5) critiqued drafts, and the authors decided which findings to apply. AI tools helped find related work and check citation metadata; every reference was verified by hand against its publisher or arXiv record. The authors take full responsibility for the content.

\bibliography{refs}
\bibliographystyle{mlsys2025}

\onecolumn
\appendix

\noindent The appendices give implementation details behind \S3--\S7 (Appendix~A), the evaluation methodology (Appendix~B), additional results (Appendix~C) and operating guidance (Appendix~D). Every number here comes from a recorded benchmark run or, where stated, the simulator.

\section{Design and Implementation Details}

\subsection{Claim Path, Durability and Restart Recovery}

\para{Order of operations.} For each queued job the dispatcher (1) claims the job's full pod set in the in-memory tracker under its write lock, which validates the transition, checks per-caller GPU and pod limits and records the lease; (2) prepares the lease on the pods (and any per-job storage); (3) commits the job's \code{QUEUED}$\rightarrow$\code{RUNNING} transition with its pod bindings and acquisition path to the durable store, conditional on the job row's version, so a stale or concurrent writer cannot move it; (4) marks the pods running in the pool, persisting owner and timing annotations in the pod PATCH the broker already issues; and (5) launches the job through the executor and pod agent. If (2) fails the pods are recycled and the job stays queued; if (3) fails the pods are released back to the tracker. Step (4) is best-effort; a launch failure after (3) is an attempt failure, retried or terminal. No user code runs before the durable transition commits.

\para{Attempts.} A job is the caller's intent; each execution is an attempt identified to the pod agent as \code{<job-id>:<retry-count>}, so a re-queued attempt can never be confused with an earlier one at the agent. Automatic retries are bounded by \code{max\_retries}.

\para{Restart.} On start, before dispatching, the broker runs a synchronous reconcile that rebuilds the tracker from the Kubernetes pods \fp{} owns (labels and the GPU scheduler's card annotation give pod state, physical GPU and standby role; annotations give the last caller, release time and remaining owner hold), and from the job store. The job store is authoritative for job state; the pod annotations are authoritative for ownership. For each job recorded as \code{RUNNING}, recovery reads its pod bindings and probes the pod agents: a live attempt is re-attached; one that finished during the outage is finalized without a retry; and one whose pods or bindings are gone, that the agent does not know (never launched), or that stays unreachable after bounded retries is re-queued (\code{RUNNING}$\rightarrow$\code{QUEUED}) and its pod released. The old attempt is not fenced before the retry is dispatched: the release cannot reset an unreachable pod, so the broker deletes it and Kubernetes terminates the attempt asynchronously, and distinct attempt IDs keep the agent from confusing the two. A network partition can therefore let both attempts run briefly; partitions are outside the tested scope. A multi-pod job is treated as a unit: if any rank is unreachable, the set stays assigned until every rank is accounted for. The scale-down cooldown is rebuilt from pod creation times, so a restart does not delete warm pods the previous process would have kept, and standby promotions fail closed until a live node sweep re-verifies GPU exclusivity (A.4). In the crash benchmark (30 \code{SIGKILL}s under load; C.3) no job was lost or ran twice and all 90 owner holds were restored.

\subsection{Packages, Dependencies and Multi-Pod Jobs}

\para{Packages.} User code is a zip archive stored content-addressed, so identical uploads are deduplicated. At job start the broker either streams it to the pod agent or, with an object store configured, hands the agent a short-lived presigned URL, keeping the broker off the bulk data path. Limits: 512\,MiB compressed (configurable), 2\,GB written during extraction (a zip-bomb guard), a 5-minute push timeout, and rejection of symlinks and paths that escape the workspace. On release, the agent empties the package directory. Stored packages are retained under the object store's lifecycle rules, which should outlast the longest queue time.

\para{Incremental dependencies.} The client bootstrap installs a requirements file into the pod's interpreter (with \code{uv}, falling back to \code{pip}) and writes a completion marker named by the SHA-256 of the file into \code{site-packages}, only after the install exits cleanly. Before every install it removes all markers, so at most one marker exists: that of the last pin set installed completely. Returning to an earlier pin set therefore reinstalls it rather than trusting a stale marker. Installs add and upgrade; a package dropped from the pins stays until the pod retires, so a caller who needs a clean environment asks for a single-use pod. Owner-only reuse confines this residue to one caller. Reusable digest-pinned images (\S4) are fenced by caller, pool, queue, reuse tag and image digest, and count against the pool's ceiling and the global GPU limits but not its idle cap.

\para{Multi-pod jobs.} Rank and rendezvous variables injected by the executor cannot be overridden by user environment. A multi-pod claim is all-or-nothing over idle pods and promotable standbys on distinct cards; if package distribution fails on any pod the launch fails, and release resets every pod of the set. The dispatcher does not reserve pods for a job it cannot place yet: it skips the job so smaller jobs behind it proceed, and the scaler counts the job's $N$ pods as demand.

\subsection{Reuse Boundaries and Reset Scope}

\para{What crosses reuse.} On every release the pod agent kills the job's whole process tree and verifies against \code{/proc} that it is gone (waiting a bounded time for tasks in uninterruptible sleep), then empties the package root, \code{/dev/shm}, volumes marked for reset and the per-attempt GPU accounting files. A reset that cannot confirm all of this reports unclean and the broker recycles the pod.

\para{Boundaries.} Reuse is scoped by caller and by a caller-supplied reuse tag (\code{lease\_tag}) that names a session. A used idle pod is claimable only by its last caller (owner-only reuse). Jobs sharing the same caller, queue and reuse tag reuse the warm pod; when a caller switches to a new session or a hold expires, the prior pod is retired, so state from one session reaches at most one job of the same caller's next session. Operators can configure the tag as a soft routing hint that prefers affinity without retiring pods.

\para{Owner holds.} After release a pod is held for its last caller for \code{caller\_hold\_seconds} (inheriting the idle timeout when unset). The caller and release time are persisted as pod annotations in requests the broker already makes, so holds survive restarts. A hold is broken early only when a queued job is blocked at the pool ceiling, and only on pods older than a 60\,s minimum tenure, oldest hold first; holds whose owner has queued jobs are spared. A broken hold retires its pod; it is never handed to another caller.

\para{Isolation.} Owner-only reuse, retirement at session boundaries, verified process and scratch-space resets, and environment-variable filtering isolate callers from each other and bound what crosses between a caller's sessions; they compose with VM-level sandboxes where kernel-level isolation is required.

\subsection{Standby Mechanics}

\para{Shares and memory cap.} On H200 the GPU-sharing scheduler (HAMi with Volcano) places one active pod and \code{standbys\_per\_active} standbys on fractional shares of one card, pinned to it by GPU UUID, and caps each pod's memory at its share through a CUDA interception library. A standby runs only the pod agent: no user code, no CUDA context, no allocation. When a job starts on the active pod, the agent sets \code{CUDA\_DEVICE\_MEMORY\_LIMIT\_<i>} to the card's full framebuffer in the job \emph{process} environment, which wins over the values injected at admission, and points \code{CUDA\_DEVICE\_MEMORY\_SHARED\_CACHE} at a fresh per-attempt file, so a reused pod does not inherit an earlier limit. The pod spec is never changed, so standbys keep their fractional cap.

\para{Gate and exclusivity.} At most one \fp{} pod per card serves at a time, so the gate promotes a standby once the previous job has exited. A pod reset for reuse is released only after its process tree is verified dead; the driver frees a context's memory when its processes exit, so gating on process death is more reliable than polling device memory. A pod marked for retirement is deleted off the critical path while the promoted standby serves. A periodic sweep checks that every share on a card belongs to \fp{}'s active--standby group: a card missing from the usage report counts as non-exclusive, a failed read changes nothing that tick, and after a restart promotions stay disabled until the sweep has verified the card. The group counts as one physical GPU in pool ceilings and global budgets.

\subsection{Scaler}

Each reconcile pass counts processors (GPUs or cores) rounded to whole pods; an active--standby group counts once and standbys are never supply. $\mathit{queued}$ sums the pods of every queued warm job, including jobs waiting on their caller's quota, so such jobs can drive creation up to the ceiling. A held pod matched by its owner's own queued jobs counts as $\mathit{idle}$ for that pass, as supply for that owner only. With $\mathit{demand}$, $\mathit{eligible}$ and $\mathit{physical}$ as in \S7 ($\mathit{idle}$ excludes other held pods; $\mathit{stale}$ covers old-template and single-use pods, which occupy capacity but cannot serve warm jobs), the pass computes
\begin{align*}
\mathit{floorDeficit} &= \max\bigl(0,\ \mathit{floor} - (\mathit{physical} - \mathit{singleUse})\bigr), \qquad \mathit{demandDeficit} = \max(0,\ \mathit{demand} - \mathit{eligible}),\\
\mathit{add} &= \min\bigl(\max(\mathit{floorDeficit}, \mathit{demandDeficit}),\ \mathit{ceiling} - \mathit{physical},\ \mathit{step},\ \mathit{headroom}\bigr).
\end{align*}
Single-use pods occupy the ceiling but cannot satisfy the floor. When nothing is created, scale-down considers only idle pods whose keep-alive has expired:
\begin{align*}
\mathit{idleCap} &= \min\bigl(\mathit{maxIdle},\ \mathit{ceiling} - (\mathit{physical} - \mathit{idle})\bigr),\\
\mathit{retain} &= \max\bigl(\mathit{idleCap},\ \mathit{floor} - (\mathit{held} + \mathit{assigned/running}),\ \mathit{demand} - (\mathit{starting}+\mathit{assigned/running})\bigr),\\
\mathit{remove} &= \max\bigl(0,\ \min(\mathit{idle} - \mathit{retain},\ \mathit{expiredIdle},\ \mathit{physical} - \mathit{floor})\bigr).
\end{align*}
Here $\mathit{assigned/running}$ is $\mathit{assigned}+\mathit{running}$. Because $\mathit{retain}$ keeps $\mathit{floor} - (\mathit{held} + \mathit{assigned/running})$ idle pods, scale-down never takes warm pods below the floor even when stale pods inflate $\mathit{physical}$. Global per-accelerator budgets are reserved in an in-memory ledger before the Kubernetes create call. Defaults: reconcile every 30\,s, hold-break minimum tenure 60\,s.

\subsection{Caller API and Errors}

The REST API decodes requests with unknown fields rejected, so a typo'd option fails loudly instead of being ignored. Job states are \code{PENDING}, \code{QUEUED}, \code{RUNNING}, \code{SUCCEEDED}, \code{FAILED} and \code{CANCELLED}. A terminal job carries an error class (e.g.\ \code{OOM}, \code{NonZeroExit}), the exit code of the final attempt's rank-0 process (absent, rather than 0, when no process ever exited), its retry count, and the acquisition path of its pods (warm, recycled, standby, new, custom image) with the reason a warmer path was not used. Callers authenticate with cluster-issued or federated identity tokens (cached validation: 6.6\,$\mu$s).

\subsection{Borrowed Capacity}

A capacity controller publishes, per GPU model, the reserved capacity that is currently unused. \fp{} polls it (default 10\,s) and may place warm or burst pods in that headroom up to the published ceiling. When an owner reclaims, the controller lowers the ceiling; \fp{} records the reclaiming owner's quota demand, as reported by the controller, as demand and preempts borrowed pods (default every 15\,s), releasing one GPU per GPU the owner needs. On kwok with a stub controller (C.3), borrowing took 11.8\,s, release to a lowered ceiling 30.0\,s, and owners got 80 of 80 pods with exactly one borrowed GPU released per owner GPU.

\section{Evaluation Methodology}

\subsection{Venues}

Headline benchmarks run in dedicated benchmark pools on production nodes. \textbf{Production H200}: one pinned 8\x{}H200 node per campaign, 1-GPU pods, a cached 12.2\,GB base image and a two-GPU cap (\S9.1 H200 columns, \S9.3, paper Table~1). \textbf{Production CPU}: 4\,vCPU/8\,GiB pods, a cached 4.4\,GiB image and a 16-pod cap, beside steady tenant load (\S9.1 CPU columns). \S8 uses production fleet telemetry: broker metrics, per-job records and pool gauges. Mechanism costs (\S9.4--9.6, C.2--C.5) run on a \textbf{local cluster}, single-host Kubernetes (k3d) on a shared 8-vCPU VM with runs serialized under a host lock and a quiet-load guard. Tracker, dispatch, catalog and authentication microbenchmarks (C.4) run on a \textbf{quiet host}, a 64-vCPU cloud VM (AMD EPYC 7763 / Xeon 8370C) with stock MySQL 8.0.33 and PostgreSQL 16.15. API load to 20k pods and borrowing (A.7) use \textbf{kwok}: a real API server, etcd and scheduler with simulated nodes and a MySQL store. Frontiers and sweeps (\S9.2, \S9.4, C.1) use the \textbf{simulator}, which runs \fp{}'s control-plane code (pool, tracker, scaler, dispatcher) in virtual time with calibrated venue models.

\subsection{Workloads}

Workloads follow \S9 and are parameterized from production job shapes; job bodies are fixed-duration payloads, so differences between arms come from acquisition, not work. Per repeat and arm, W-RL submits 44 jobs; W-Agent submits 19 jobs with 2 agents (R3), 37 with 4 agents on CPU and 36 in the cap sweep; and the standby benchmark runs one caller's 30 sequential W-Agent jobs, each with a fresh reuse tag. Makespan runs from a repeat's first submission to its last completion.

\subsection{Arms}

\textbf{Reuse On}: a stable reuse tag per caller, so each caller's steps share its warm pod. \textbf{Reuse Off}: a fresh reuse tag per job with the tag as a retiring boundary: a job may run once on the previous job's pod, which then retires, or land on an unheld idle, recycled or standby pod; this arm ablates owner reuse. \textbf{Fresh pod per job}: each job asks for the pool's own image with reuse disabled, so \fp{} creates a single-use pod for each job and deletes it at release, the closest analogue of a Kubernetes Job.

\subsection{Statistics}

Arms are interleaved per repeat in a Williams crossover design to balance carry-over (e.g.\ warm pods left by the previous arm). Percentiles pool all jobs of an arm with linear interpolation. Paired effects are computed per repeat and summarized with 95\% bootstrap CIs over repeats (seed 2027, 2,000 resamples); the start-latency effect is the median of per-repeat p50 differences, and the makespan effect the median paired difference.

\para{Simulator calibration.} The simulator's venue constants (per-phase pod start, termination and dispatch costs) are fitted on train campaigns and validated on held-out campaigns registered in advance; a fit is adopted only if it does not worsen any hold-out set. Under the adopted configuration, 46 of 55 held-out metrics (84\%) fall within band, 52 of 65 (80\%) including the H200 run (R3), and 132 of 145 of a second hold-out set fall inside their 95\% prediction intervals. Prediction intervals are $\times$0.69--1.44 for queue medians and $\times$0.76--1.31 for makespan and throughput. Rankings of arms are preserved in every hold-out set and all effect signs are correct. We use the simulator for shapes, orderings and crossovers, never for headline numbers.

\section{Additional Results}

\subsection{Held Capacity, Job Length and Think Time}

\para{Simulated frontier (production CPU pod start).} Held processor-time relative to an always-on pool sized to peak, median of 3 seeds, with $\Delta$p50 start against that pool ($\Delta$p50 not recorded at $\tau=300$\,s):

\begin{table}[H]
\centering\small
\begin{tabular}{@{}lcccccc@{}}
\toprule
Pattern & $\tau=15$\,s & $\tau=60$\,s & $\tau=120$\,s & $\tau=300$\,s & $\tau=900$\,s & Fresh pod per job \\
\midrule
Steady & 0.435 (+0.38\,s) & 0.571 (+0.08\,s) & 0.679 (+0.01\,s) & 0.916 & 1.543 ($-$0.02\,s) & 1.347 (+22.8\,s) \\
Bursty & 0.371 (+0.64\,s) & 0.476 (+0.27\,s) & 0.615 (+0.13\,s) & 0.849 & 1.484 (+0.09\,s) & 0.646 (+23.0\,s) \\
Diurnal & 0.486 (+0.37\,s) & 0.606 (+0.01\,s) & 0.659 (+0.03\,s) & 0.860 & 1.124 ($-$0.01\,s) & 1.543 (+22.9\,s) \\
\bottomrule
\end{tabular}
\end{table}

With 22\,s pod starts the scaler creates more pods than the peak concurrency during bursts (11--16 against 5--7), because queued calls trigger creates while released pods free up seconds later; a 900\,s keep-alive keeps that overshoot through the whole tail. A fresh pod per job is not the cost floor at production pod-start times.

\para{Measured frontier (local cluster).} 2,300 jobs, 5 paired repeats: a floor of 1, idle cap of 2 and 60\,s keep-alive held 0.52\x{} [0.52, 0.53] the always-on pool for +28\,ms at the median; a bare 60\,s keep-alive 0.24\x{} for +46\,ms; a 900\,s keep-alive (no idle cap) 1.15\x{} [1.09, 1.18] for +57\,ms.

\para{Simulated frontier with a floor.} With production's 22\,s pod start, a floor of 4 with an idle cap of 2 buys back the bursty tail: p90 3.39\,s at 0.78\x{} the always-on pool.

\para{H200 GPU time.} From per-minute pool gauges during the R3 run: warm reuse held 2,340 GPU-s for 190 jobs (12.3 per job; 87\% idle, almost all owner-held between agent steps); Reuse Off held 2,040 GPU-s (10.7 per job; 35\% creating pods). Standbys add no physical GPUs.

\begin{figure}[h]
\centering
\includegraphics[width=\textwidth]{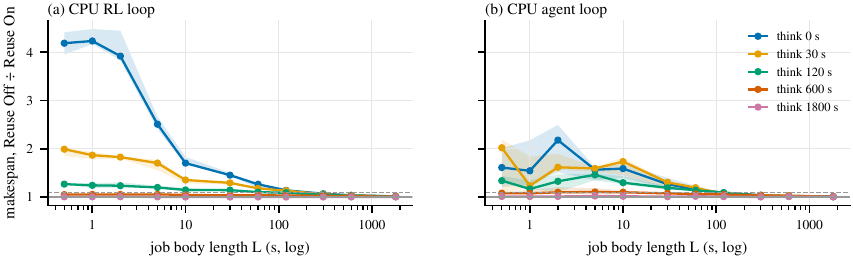}
\caption{Loop speedup of Reuse On over Reuse Off against job body length and think time (simulator, production CPU calibration; bands span 3 replicates). The gain falls below 1.1\x{} at about 196\,s (RL) and 104\,s (agent) bodies with no think time, and vanishes when think time exceeds the 900\,s keep-alive. At 1--2\,s bodies the sweep gives 4.1--4.3\x{} for RL (measured 4.5\x) and 1.6--2.0\x{} for agents (measured 2.2\x).}
\end{figure}

\subsection{Isolation Costs (local cluster)}

\para{Pod-agent reset} (130 trials on one pod, owner-only reuse on): clean 9.3\,ms; 10,000 scratch files 187\,ms; 8\,GiB of \code{/dev/shm} 1.35\,s; leftover processes about 110\,ms (one re-check after \code{SIGKILL}); worst case (all three) 1.67\,s. Injected unremovable residue: the agent reported unclean, the pod was recycled 17\,ms after the job finished, and the next job ran on a new pod (5/5, 0 failed jobs). Three concurrent callers: 0 cross-caller reuse, Jain goodput fairness 0.976.

\subsection{Failure, Rollout and Borrowing}

\para{Crash.} 10 repeats (5 agent with 3 callers, 5 RL), 300\,s each, \code{SIGKILL} at 60, 150 and 240\,s: 30 kills. 0 lost and 0 duplicate jobs; held 90/90, unheld idle 92/92 and busy 27/27 pods kept; process start to dispatching 0.045\,s p50 (0.068\,s p90); first claim 3.0\,s after start; kill to first API response through the Service 20\,s (kubelet back-off and readiness probe).

\para{Template rollout under saturated load.} An 8-pod pool (floor 4, idle cap 6, ceiling 8, drain budget 1), 12 callers with no think time, a template edit: converged in 324\,s [286, 359]; all 12 callers served in every run; longest wait 181--242\,s; 0 overtakes of the longest waiter; 0 failed jobs; at most one pod retiring and never more than 8 pods; $\Delta$p50 +0.03\,s.

\para{Borrowing (kwok).} 10 ceiling cycles and 10 owner cycles: time to borrow 11.8\,s [10.5, 13.2]; release to a lowered ceiling 30.0\,s; owners placed 80/80 pods within 240\,s; 1.0 borrowed GPU released per owner GPU in all cycles; 0 over-placement.

\subsection{Control-Plane Scale}

\para{Warm start.} With an idle H200 pod, a warm start took 1.85\,s under the 3\,s dispatch tick (1.27\,s of it waiting for the tick) and takes 0.59\,s with event-driven dispatch: accept 0.04, dispatch 0.01, hand-off 0.39 and first instruction 0.12\,s.

At 100,000 pods the tracker sustains 191--219k claims/s with one claimer and 109--120k with 32; a claim holds the write lock 1.5--2.6\,$\mu$s at the median ($\leq$8\,$\mu$s p99); lock-wait p99 is 0.66\,ms at 32 claimers; a reconcile pass takes at most 0.56\,s of its 30\,s interval; the index costs about 1\,KB per pod. Durable MySQL dispatches 33 jobs/s with one goroutine and 483 with 64 (21\,ms per commit); PostgreSQL 184 and 2,365 (3.7\,ms); with the database on another host, 244 and 1,505 at 64 goroutines. The 5,025 jobs/s we measured on in-memory SQLite is an engine bound, not a production figure. Catalog load: 1,000 pool definitions in 347\,ms. Authentication: cached service-account token 6.6\,$\mu$s, uncached with local keys 47\,$\mu$s. In production the broker sends only a few requests per second to each API server without throttling; on kwok to 20,000 pods across 3,282 nodes, API priority-and-fairness never throttled (p99 wait $\leq$0.13\,s), and \fp{}'s own serial pod creation and 100-QPS client limiter bound first.

\subsection{Late Binding Details}

The simulator behind paper Figure~5 runs \fp{}'s control plane with a 22.1\,s cached cold start, a 64-pod budget, a 900\,s keep-alive, Zipf ($s=1$) environment popularity and 10 paired seeds (3,360 cells). With per-image cold starts of 300\,s, per-environment pools hold 5.96\x{} the always-on pool on RL rollouts while the shared pool stays at 1.06\x{}; with a 30\,s per-job hand-off, the shared pool holds as much as per-environment pools at $E=64$ (2.08\x{} vs 2.15\x{}).

\para{Dependencies.} First install 0.11\,s (3 pins), 2.5\,s (23 pins), 4.3\,s (CPU PyTorch); repeated pins 0 (marker hit); one changed pin 0.03--0.06\,s.

\para{Ray.} Submit to first task on a warm $N$-pod Ray cluster formed at job start: 8.6\,s ($N$=2), 9.9\,s ($N$=4), 13.1\,s ($N$=8), against 14.4--19.6\,s cold. Warm pods save the pod start but not Ray's own bootstrap (driver start 2.0--2.6\,s, \code{ray.init} 0.4\,s, first task 1.1\,s after the last node registers). Multi-pod warm claims for torchrun-style jobs take 0.27--0.45\,s for $N$=2--8, with 0 partial claims in 50 jobs; an 8-pod job behind a 3-pod holder held no pods while waiting and claimed all 8 at once.

\subsection{H200: Fresh Pod per Job and Reuse Off}

\para{Fresh pod per job (cap sweep).} Paired repeat-makespan differences, fresh pod minus warm reuse, under the two-GPU cap: +1,438\,s [1,314, 1,769] (agent) and +1,572\,s [1,423, 2,097] (RL), 10 paired repeats each.

\para{Acquisition paths (R3).} Warm / standby / recycled / new pods: Reuse On 171 / 11 / 4 / 4; Reuse Off 33 / 66 / 71 / 20.

\section{Operating Guidance}

\noindent We follow five operating rules:
\begin{itemize}
\item \textbf{Judge warm pools by tails, loops and held capacity.} Our H200 medians differ by 0.4\,s while p90 differs by 55\,s and the loop by 3.6\x; report p90, loop makespan and held GPU time alongside the median.
\item \textbf{Size keep-alives to the loop's pauses.} A keep-alive just above think time bridges pauses; a long one keeps every burst's overshoot. Owner holds partition idle pods by caller, so size idle caps by active callers.
\item \textbf{Make each isolation boundary a per-tenant choice.} A hard per-tag boundary kept half of jobs warm and a routing hint 99\%; choose per tenant.
\item \textbf{Build for a caller that is a program.} Reject unknown fields and return typed errors to stop retry storms; record how each pod was obtained to tell cold-start regressions from quota exhaustion.
\item \textbf{Test the control plane as hard as the data plane.} Once pods are warm, the dispatch wake-up and the durable commit dominate a warm start (C.4); fault-inject broker restarts, template rollouts and capacity reclaims.
\end{itemize}

\end{document}